\documentclass[12pt]{article}
\usepackage{amsmath}
\usepackage{graphicx,psfrag,epsf,xcolor}
\usepackage{enumerate}
\usepackage{natbib}
\usepackage{url} 
\usepackage{amsfonts}

\usepackage{lineno}

\usepackage{amssymb}
\usepackage{mathtools}
\usepackage{algorithm}
\usepackage[noend]{algpseudocode}
\usepackage{subcaption}

\usepackage{amsthm}

\newtheorem{theorem}{Theorem}

\newcommand{\blind}{0}

\begin{document}

\def\spacingset#1{\renewcommand{\baselinestretch}%
{#1}\small\normalsize} \spacingset{1}


\if0\blind
{
  \title{\bf Nonparametric Estimation of Conditional Copula using Smoothed Checkerboard Bernstein Sieves}
  \author{Lu Lu\thanks{
    Corresponding author Email: lulu1210143@gmail.com}\hspace{.2cm}
    and 
    Sujit K. Ghosh \\
    Department of Statistics, North Carolina State University}
  \maketitle
} \fi

\if1\blind
{
  \bigskip
  \bigskip
  \bigskip
  \begin{center}
    {\LARGE\bf Nonparametric Estimation of Conditional Copula using Smoothed Checkerboard Bernstein Sieves}
\end{center}
  \medskip
} \fi

\bigskip
\begin{abstract}

Conditional copulas are useful tools for modeling the dependence between multiple response variables that may vary with a given set of predictor variables. Conditional dependence measures such as conditional Kendall's tau and Spearman's rho that can be expressed as functionals of the conditional copula are often used to evaluate the strength of dependence conditioning on the covariates. In general, semiparametric estimation methods of conditional copulas rely on an assumed parametric copula family where the copula parameter is assumed to be a function of the covariates. The functional relationship can be estimated nonparametrically using different techniques but it is required to choose an appropriate copula model from various candidate families. In this paper, by employing the empirical checkerboard Bernstein copula (ECBC) estimator we propose a fully nonparametric approach for estimating conditional copulas, which doesn't require any selection of parametric copula models. Closed-form estimates of the conditional dependence measures are derived directly from the proposed ECBC-based conditional copula estimator. We provide the large-sample consistency of the proposed estimator as well as the estimates of conditional dependence measures. The finite-sample performance of the proposed estimator and comparison with semiparametric methods are investigated through simulation studies. An application to real case studies is also provided.

\end{abstract}

\noindent%
{\it Keywords:} empirical checkerboard Bernstein copula (ECBC), conditional dependence measures, covariates
\vfill

\newpage
\spacingset{1.45} 
\section{Introduction}
\label{sec:intro}

Copulas have found many applications in the field of finance, insurance, system reliability, etc., owing to its utility in modeling the dependence among variables (see, e.g., \citet{nelsen2007introduction}, \citet{jaworski2010copula} and \citet{joe2014dependence} for details about copulas and their applications).
In some situations, the dependence structure between variables can be influenced by a set of covariates and it is thereby of interest to understand how such dependence changes with the values of covariates. For instance, it is well known that the life expectancy at birth of males and females in a country are often highly dependent due to
shared economic or environmental factors and it is possible that the strength of dependence relies on these factors. When the covariate is binary or discrete-valued with few levels, one can estimate a copula for each given level of the discrete-valued covariate separately. In constrast, the influence of continuous-value covariate on the dependence structure should be formulated in a functional way, and
this is where conditional copulas (\citet{patton2006estimation}; \citet{patton2006modelling}) along with the corresponding conditional versions of dependence measures come into play.

Suppose we are interested in the dependence among the components of a random vector $\mathbf{Y} = (Y_1, Y_2, ..., Y_d)$, given covariates $\mathbf{X} = (X_1, X_2, ..., X_p)$. The conditional joint and marginal distribution of $\mathbf{Y}$ given $\mathbf{X} = \mathbf{x}$ can be denoted as 
\begin{equation}
F_{\mathbf{x}}(\mathbf{y}) \equiv F_{\mathbf{x}}(y_1, y_2, ... y_d ) = P (Y_1 \leq y_1, Y_2 \leq y_2, ..., Y_d \leq y_d \;| \; \mathbf{X} = \mathbf{x}),
\end{equation}
and 
\begin{equation}
F_{j \mathbf{x}}(y_j ) = P (Y_j \leq y_j \;| \; \mathbf{X} = \mathbf{x}) \; \; j = 1, ..., d.  
\end{equation}
If $F_{1 \mathbf{x}}, F_{2 \mathbf{x}}, ..., F_{d \mathbf{x}}$ are continuous, then by an extension of the well-known Sklar's theorem (\citet{sklar1959fonctions}) for conditional distributions (e.g. see \citet{patton2006modelling}), there exists a unique copula $C_{\mathbf{x}}$ such that 
\begin{equation}
F_{\mathbf{x}}(\mathbf{y}) = C_{\mathbf{x}} (F_{1 \mathbf{x}}(y_1), F_{2 \mathbf{x}}(y_2), ..., F_{d \mathbf{x}}(y_d)) \; \; \forall y \in \mathbb{R}^d, \forall y \in \mathbb{R}^d,
\end{equation}
and the function $C_{\mathbf{x}}$ is called a conditional copula, which captures the conditional dependence structure of $\mathbf{Y}$ given $\mathbf{X} = \mathbf{x}$. The focus of this paper is modeling continuous-valued responses and covariates. Thus, in what follows, we assume that the conditional marginal CDFs $F_{1 \mathbf{x}}, j = 1, \ldots, d$ and the CDFs of each response and covariate are absolutely continuous.

The literature contains a variety of parametric families for modeling copulas. Some commonly used copula families are Archimedean copulas, elliptical copulas, etc.; see \citet{vzevzula2009multivariate} and \citet{joe2014dependence}, etc. Assuming that the conditional copula belongs to a parametric copula family where the copula parameter is a function of the covariate(s), there has been previous work addressing the estimation of conditional copula in a semiparametric setting. In regard to frequentist methods based on an assumed parametric class, \citet{acar2011dependence} propose to estimate the functional relationship between the copula parameter and the covariate nonparametrically by using local likelihood approach. But they assume known marginals and the maximization is conducted for a fixed value of the covariate, i.e., With the intention of identifying the entire function between the copula parameter and the covariate, it is necessary to solve the maximization problem for a sufficiently large grid of values within the range of the covariate. \citet{abegaz2012semiparametric} extend the work to a more general setting of unknown marginals and apply a two-stage technique that has been widely adopted in copula estimation: in the first stage, the nonparametric estimates of conditional marginals are obtained using kernel-based method and by plugging in theses estimates, the functional link is estimated by maximizing the pseudo log-likelihood in the second stage. As alternative estimation methods for the function relationship, \citet{vatter2015generalized} develop generalized additive models for conditional dependence structures, and \citet{fermanian2018single} introduce so-called single-index copulas, etc. In the Bayesian framework, inference for bivariate conditional copula models have been constructed in \citet{craiu2012mixed}, \citet{sabeti2014additive} and \citet{levi2018bayesian}, among others.

However, the misspecification of copula family could lead to severely biased estimation even though a sophisticated and flexible parametric model is employed (e.g. see \citet{geerdens2018conditional}), so it is required to select an appropriate copula model from a large number of candidate families. In order to do so, many copula selection techniques have been proposed either in either frequentist or Bayesian setting, e.g., \citet{acar2011dependence} select the copula family based on cross-validated prediction errors, while the deviance information criterion (DIC) is utilized for the choice of copula in \citet{craiu2012mixed}.

Acknowledging the limitations of parametric copula models as mentioned above, fully nonparametric approaches have also been proposed for conditional copula estimation. 
\citet{gijbels2011conditional} suggests the empirical estimators for conditional copulas where the weights are smoothed over the covariate space through kernel-based methods. They further derive nonparametric estimates for the conditional dependence measures including conditional Kendall's tau and conditional Spearman's rho. Since the bandwidth selection is very crucial for any of the smoothing methods, they also develop an algorithm for selecting the bandwidths. The asymptotic properties of the estimators together with conditional dependence measure estimates are established in \citet{veraverbeke2011estimation}. \citet{gijbels2012multivariate} further consider more complex covariates like multivariate covariates and box-type conditioning events are studied in \citet{derumigny2020conditional}. On the other hand, there has been recent work on Bayesian nonparametric estimation of conditional copula. \citet{leisen2017bayesian} introduce the effect of a covariate to the Bayesian infinite mixture models proposed by \cite{wu2014bayesian}. However, large-sample asymptotic properties of the Bayesian models have been almost unexplored and still remains an area of open work.

In this paper, we focus on the nonparametric estimation of conditional copulas and have realized that it can be done in a relatively easy way by employing the empirical checkerboard Bernstein copula (ECBC) estimator. When the covariates are continuous-valued, the main idea of extending the copula models to include covariates is to first estimate the full copula of responses along with covariates and then take partial derivatives to obtain the conditional distribution of responses given the covariates.
As a fully nonparametric approach, it is not required to make any selection of the proper copula family, which is a key step in semiparametric methods to avoid the adverse consequence of model misspecification. Compared to the kernel-based empirical estimators, the selection of bandwidths is unnecessary either, making it easy to implement in practice. The proposed ECBC-based conditional copula estimator immediately leads to nonparametric estimates of the conditional dependence measures, which can be expressed in a very neat form under matrix operations.
The large-sample consistency of the proposed estimator is also provided in the paper.

\section{Models for Conditional Copula}
\label{sec:meth}

In the following, we focus on the bivariate conditional copula of $(Y_1, Y_2)$ with a single covariate $X$ for simplicity. Notice that the extension to more than two
dimensions and multiple covariates is straightforward. 

Suppose we have i.i.d. samples  $(Y_{i1}, Y_{i2}, X_{i}), i = 1, ..., n$, where $(Y_{i1}, Y_{i2}), i = 1, ..., n$ are i.i.d observations of the random vector $(Y_1, Y_2)$ of which the conditional dependence structure is of our interest. $X_{i}, i = 1, ..., n$ are i.i.d. observations of the covariate $X$. We assume all components of $(Y_1, Y_2, X)$ are continuous-valued random variables with absolutely continuous marginal distributions and the conditional marginal distributions of $Y_1$ and $Y_2$ given $X = x$ are also absolutely continuous. The goal is to estimate the conditional copula $C_{x}$ from a random sample of i.i.d. observations $(Y_{i1}, Y_{i2}, X_{i}), i = 1, ..., n$.

As suggested by \citet{gijbels2011conditional}, it is often favorable to remove the effect of the covariate on the marginal distributions before estimating $C_{x}$. In order to do that, we can transform the original observations $(Y_{i1}, Y_{i2})$ to marginally uniformly distributed (unobserved) samples 
\begin{equation}
(U_{i1}, U_{i2}) \equiv (F_{1 X_i}(Y_{i1}), F_{2 X_i}(Y_{i2})), \; \; i=1, ..., n,
\end{equation}
which can be estimated by pseudo-observations
\begin{equation}
\label{eq:pseudo}
(\hat{U}_{i1}, \hat{U}_{i2}) \equiv (\hat{F}_{1 X_i}(Y_{i1}), \hat{F}_{2 X_i}(Y_{i2})), \; \; i=1, ..., n,
\end{equation}
where $ \hat{F}_{1 x}$ and $\hat{F}_{2 x}$ are the estimated conditional marginal distributions. 

Motivated by \citet{janssen2016bernstein} who apply the empirical Bernstein estimator of bivariate copula derivative to conditional distribution estimation with a single covariate, we are able to use the proposed multivariate copula estimator ECBC to estimate the conditional marginal distributions of $Y_1$ and $Y_2$ given $X = x$, respectively.
Specifically, for $j \in \{1, 2\}$, we have i.i.d samples $(Y_{ij}, X_{i}), i = 1, ..., n$ and the corresponding pseudo-observations $(\hat{W}_{ij}, \hat{V}_{i} ) =$ $(F_{nY_j}(Y_{ij}), F_{nX}(X_{i})), i =1, ..., n$, where $F_{nY_j}$ and $F_{nX}$ are the modified empirical estimation of the (unconditional) marginal distributions $F_{Y_j}$ and $F_{X}$, respectively, e.g., $ F_{n X}(x) =  \frac{1}{n+1} \sum_{i=1}^{n} \mathbb{I}(X_{i} \leq x)$. These pseudo-observations can be then treated as samples from a $2$-dimensional copula $C_j$, which can be estimated by the ECBC copula estimator as follows
\begin{equation}
C^{\#}_j (w_j, v) = \sum_{h = 0}^{g_j} \sum_{k = 0}^{m_j} \tilde{\theta}_{h,k}    {g_j \choose h } w_j^{h} (1 - w_j)^{g_j - h} {m_j \choose k } v^{k} (1 - v)^{m_j - k},
\end{equation}
where 
\begin{equation}
\tilde{\theta}_{h,k}  = C^{\#}_{jn}(\frac{h}{g_j}, \frac{k}{m_j}),
\end{equation}
and $C^{\#}_{jn}$ is the empirical checkerboard copula. Then the partial derivative $C^{(1)}_j$ of $C_j$ with respect to $v$ can be estimated by using
\begin{equation}
\begin{aligned}
\label{eq:partial}
C^{\# (1)}_j (w_j, v) &\equiv \frac{\partial C^{\#}_j (w_j, v)}{\partial v} \\
&= \sum_{h = 0}^{g_j} \sum_{k = 0}^{m_j-1}  \tilde{\lambda}_{h,k}    {g_j \choose h } w_j^{h_j} (1 - w_j)^{g_j - h}         m_j {m_j-1 \choose k } v^{k} (1 - v)^{m_j - k -1},
\end{aligned}
\end{equation}
where
\begin{equation}
\begin{aligned}
\tilde{\lambda}_{h,k}  &=  \tilde{\theta}_{h,k+1} -  \tilde{\theta}_{h,k}.
\end{aligned}
\end{equation}
Notice that the following relationship holds between the conditional marginal distribution function of
$Y_j$ given $X = x$ and the partial derivative $C^{(1)}_j (w_j, v)$
\begin{equation}
\begin{aligned}
F_{j x} (y_j) = P(Y_j \leq y_j \; | \; X = x) = C^{ (1)}_j (F_{Y_j}(y_j), F_{X}(x)).
\end{aligned}
\end{equation}
Thus, we can estimate the conditional marginal distributions using 
\begin{equation}
\begin{aligned}
\hat{F}_{j x} (y_j)  = C^{\# (1)}_j (F_{n Y_j}(y_j), F_{n X}(x))
\end{aligned}
\end{equation}
for $j = 1, 2$, and then the corresponding pseudo-observations $(\hat{U}_{i1}, \hat{U}_{i2}), i=1,... , n$ of the conditional copula $C_{x}$ adjusted for the effect of the covariate on the marginal distributions can be estimated as given in (\ref{eq:pseudo}).

Now we can use the covariate-adjusted pseudo-observations $(\hat{U}_{i1}, \hat{U}_{i2}), i = 1, ..., n$ along with the pseudo-observations of the covariate $\hat{V}_{i}, i = 1, ..., n$ to estimate a $3$-dimensional copula $C(u_1, u_2, v)$ again using ECBC and denote it as $C^{\#}(u_1, u_2, v)$. Similar to (\ref{eq:partial}), it is easy to obtain the partial derivative $C^{\# (1)}$ of $C^{\#}$ with respect to $v$, which is denoted as
\begin{equation}
\begin{aligned}
C^{\# (1)} (u_1, u_2 | v) &\equiv \frac{\partial C^{\#} (u_1, u_2, v)}{\partial v } \\
&= \sum_{h_1 = 0}^{l_1} \sum_{h_2 = 0}^{l_2} \sum_{k = 0}^{m-1} \tilde{\gamma}_{h_1, h_2,k}   m {m-1 \choose k } v^{k} (1 - v)^{m - k -1}   \prod_{s=1}^2 {l_s \choose h_s } u_s^{h_s} (1 - u_s)^{l_s - h_s}   ,
\end{aligned}
\end{equation}
where
\begin{equation}
\begin{aligned}
\tilde{\gamma}_{h_1, h_2, k}  &= \tilde{\theta}_{h_1, h_2, k+1} - \tilde{\theta}_{h_1, h_2, k}.
\end{aligned}
\end{equation}
Notice that we can use $C^{\# (1)} (u_1, u_2 | F_{n X}(x))$ as an estimate of the conditional copula $C_{x}$, however, $C^{\# (1)} (u_1, u_2 | v)$ is itself a valid bivariate copula for any value of $v \in [0,1]$ only asymptotically. This is because the conditional marginal distributions of $C^{\# (1)} (u_1, u_2 | v)$ are not necessarily uniform distributions for finite samples. Aiming to obtain a more accurate estimate of the conditional copula for small samples,  
we consider the conditional marginal distributions of $C^{\# (1)} (u_1, u_2 | v)$ given as
\begin{equation}
\begin{aligned}
F_1 (u_1 | v) &\equiv C^{\# (1)} (u_1, 1 | v) \\
&= \sum_{h_1 = 0}^{l_1} \sum_{k = 0}^{m-1} \tilde{\gamma}_{h_1, l_2, k}  m {m-1 \choose k } v^{k} (1 - v)^{m - k -1} {l_1 \choose h_1 } u_1^{h_1} (1 - u_1)^{l_1 - h_1}  ,
\end{aligned}
\end{equation}
and 
\begin{equation}
\begin{aligned}
F_2 (u_2 | v) &\equiv C^{\# (1)} (1, u_2 | v) \\
&= \sum_{h_2 = 0}^{l_2} \sum_{k = 0}^{m-1} \tilde{\gamma}_{l_1, h_2, k} m {m-1 \choose k } v^{k} (1 - v)^{m - k -1}  {l_2 \choose h_2 } u_2^{h_2} (1 - u_2)^{l_2 - h_2}  ,
\end{aligned}
\end{equation}
By using Sklar's theorem, we are able to obtain a conditional copula estimator which is a genuine copula itself denoted as 
\begin{equation}
\begin{aligned}
& C^{\#} (u_1, u_2 | v) = C^{\# (1)} (F_1^{-1}(u_1 |v ), F_2^{-1}(u_2| v) | v),
\end{aligned}
\end{equation}
where $F_1^{-1}(u_1 | v)$ and $F_2^{-1}(u_2 | v)$ are the inverse functions of $F_1$ and $F_2$, respectively. It is to be noted that $C^{\#} (u_1, u_2 | v)$ is a valid copula for any value of $v \in [0,1]$, and as a result, the conditional copula $C_{x}$ can be estimated by 
\begin{equation}
\begin{aligned}
C^{\#}_{x} (u_1, u_2) = P(F_{1 x}(y_1) \leq u_1, F_{2 x}(y_2) \leq u_2 \; | \; X = x) = C^{\#} (u_1, u_2 | F_{n X}(x)).
\end{aligned}
\end{equation}

Let $\displaystyle ||g||(v)=\sup_{(u_1, u_2)\in [0, 1]^2}| g(u_1, u_2| v)|$ denote the conditional supremum norm of a conditional function $g(u_1, u_2|v)$ defined on the unit square $[0, 1]^2$ for a fixed $v$. We denote the common supremum norm as $\displaystyle ||\cdot||$. The following theorem provides the large-sample consistency of the estimator $C^{\#} (u_1, u_2 | v)$ for fixed value of $ 0 < v < 1$ using the conditional supremum norm.

\begin{theorem}
\label{thm:consist}
Assume that the underlying trivariate copula $C(u_1, u_2, v)$ is absolutely continuous and conditional copula $C_v(u_1, u_2 | v) = \frac{\partial C(u_1, u_2, v)}{\partial v}$ is Lipschitz continuous on $[0,1]^3$. Then for any fixed $0 < v < 1$, we have
\begin{equation*}
\begin{aligned}
E(||  C^{\#}  - C_v||(v) ) \stackrel{\text{a.s.}}{\rightarrow} 0 \;\; as\; \;   n \rightarrow  \infty.
\end{aligned}
\end{equation*}
where the expectation is taken with respect to the empirical prior distribution of $l_1, l_2$, and $m$ as given for ECBC.
\end{theorem}

\noindent
\textbf{Remark}: Following the hierarchical shifted Poisson distributions proposed for ECBC in {\citet{lu2023nonparametric} , the empirical prior distribution of $l_1, l_2$, and $m$ are given as
\begin{equation*}
l_j |\alpha_j \sim Poisson(n^{\alpha_j}) + 1\;\;\mbox{and}\;\;
\alpha_j \sim Unif\Big(\frac{1}{3}, \frac{2}{3}\Big) \; \;j= 1,2,
\end{equation*}
\begin{equation*}
m |\alpha \sim Poisson(n^{\alpha}) + 2\;\;\mbox{and}\;\;
\alpha \sim Unif\Big(\frac{1}{3}, \frac{2}{3}\Big). \; \;
\end{equation*}
The choice of the above priors are motivated by the asymptotic theory of empirical checkerboard copula methods \citet{janssen2014note}. The use of sample size or more generally data dependent priors have been used extensively in literature (e.g., see \citet{wasserman2000asymptotic} and \citet{parrado2012pac}) and have been shown to produce desirable asymptotic properties of the posterior distributions.

Next, by extending the dependence measures given in  \citet{schweizer1981nonparametric} to conditional versions, we are able to estimate the conditional dependence measures (e.g. conditional Spearman's rho, conditional Kendall's tau, etc.) using the estimator $C^{\# (1)} (u_1, u_2 | v)$. For instance, the estimate of conditional Kendall's tau takes the form
\begin{equation}
\begin{aligned}
& \hat{\tau}(v) = 4 \int_{0}^{1} \int_{0}^{1} C^{\# (1)} (u_1, u_2 | v) d C^{\# (1)} (u_1, u_2 | v) - 1,
\label{eq:tau0}
\end{aligned}
\end{equation}
and the estimate of conditional Spearman's rho is given as
\begin{equation}
\begin{aligned}
& \hat{\rho}(v) = 12 \int_{0}^{1} \int_{0}^{1} \big(C^{\# (1)} (u_1, u_2 | v) - F_1 (u_1 | v) F_2 (u_2 | v) \big) d F_1 (u_1 | v)  dF_2 (u_2 | v). 
\label{eq:rho0}
\end{aligned}
\end{equation}
Let us denote 
\begin{equation}
\begin{aligned}
\eta_{h_1, h_2 | v} \equiv  m \sum_{k = 0}^{m-1} \tilde{\gamma}_{h_1, h_2, k}  {m-1 \choose k } v^{k} (1 - v)^{m - k -1}, \; \; h_1 = 0, ..., l_1, h_2 = 0, ..., l_2.
\end{aligned}
\end{equation}
Then we can rewrite the estimator $C^{\# (1)} (u_1, u_2 | v)$ and its conditional marginal distributions as  
\begin{equation}
\begin{aligned}
C^{\# (1)} (u_1, u_2 | v) 
&= \sum_{h_1 = 0}^{l_1} \sum_{h_2 = 0}^{l_2} \eta_{h_1, h_2 | v}   \prod_{s=1}^2 {l_s \choose h_s } u_s^{h_s} (1 - u_s)^{l_s - h_s},
\end{aligned}
\end{equation}
\begin{equation}
\begin{aligned}
\label{eq:f1}
& F_1 (u_1 | v) = 
\sum_{h_1 = 0}^{l_1} \eta_{h_1, l_2 | v} {l_1 \choose h_1 } u_1^{h_1} (1 - u_1)^{l_1 - h_1}, 
\end{aligned}
\end{equation}
and 
\begin{equation}
\begin{aligned}
\label{eq:f2}
& F_2 (u_2 | v) = 
\sum_{h_2 = 0}^{l_2} \eta_{l_1, h_2 | v} {l_2 \choose h_2 } u_2^{h_2} (1 - u_2)^{l_2 - h_2}, 
\end{aligned}
\end{equation}
respectively. As a result, a closed-form estimate of conditional Kendall's tau takes the form 
\begin{equation}
\begin{aligned}
\hat{\tau}(v) = 4 \sum_{h_1 = 0}^{l_1} \sum_{h_2 = 0}^{l_2} \sum_{g_1 = 0}^{l_1-1} \sum_{g_2 = 0}^{l_2-1} \eta_{h_1, h_2 | v}(\eta_{g_1+1, g_2+1 | v} - \eta_{g_1+1, g_2 | v} - \eta_{g_1, g_2+1 | v} + \eta_{g_1, g_2 | v}) \\
 \prod_{s=1}^2 l_s {l_s \choose h_s } {l_s-1 \choose g_s }  B(h_s + g_s+1, 2l_s-h_s-g_s) - 1,
\label{eq:tau1}
\end{aligned}
\end{equation}
where $B$ is the beta function. Similarly, we are able to obtain a closed-form estimate of conditional Spearman's rho as
\begin{equation}
\begin{aligned}
&\hat{\rho}(v) = 12 \sum_{h_1 = 0}^{l_1} \sum_{h_2 = 0}^{l_2} \sum_{g_1 = 0}^{l_1-1} \sum_{g_2 = 0}^{l_2-1} (\eta_{h_1, h_2 | v} -       \eta_{h_1, l_2 | v}     \eta_{l_1, h_2 | v}    )(\eta_{g_1+1, l_2 | v} - \eta_{g_1, l_2 | v})( \eta_{l_1, g_2+1 | v} - \eta_{l_1, g_2 | v}) \\
& \prod_{s=1}^2 l_s {l_s \choose h_s } {l_s-1 \choose g_s }  B(h_s + g_s+1, 2l_s-h_s-g_s). 
\label{eq:rho1}
\end{aligned}
\end{equation}

For the purpose of computing the estimates of conditional dependence measures more efficiently, we apply matrix operations to the tensor products in expressions (\ref{eq:tau1}) and (\ref{eq:rho1}). For given $(h_1, h_2), h_1 = 1, \ldots, l_1, h_2 = 1, \ldots, l_2$, let us denote 
\begin{equation}
a_{h_1, g_1} = l_1 {l_1 \choose h_1 } {l_1-1 \choose g_1 }  B(h_1 + g_1+1, 2l_1-h_1-g_1), \; \; g_1 = 0, \ldots l_1-1.
\end{equation}
and 
\begin{equation}
b_{h_2, g_2} = l_2 {l_2 \choose h_2 } {l_2-1 \choose g_2 }  B(h_2 + g_2+1, 2l_2-h_2-g_2), \; \; g_2 = 0, \ldots l_2-1.
\end{equation}
Then we have $\mathbf{a}_{h_1} = (a_{h_1,0}, \ldots, a_{h_1,l_1-1})^T$ and $\mathbf{b}_{h_2} = (b_{h_2,0}, \ldots, b_{h_2,l_2-1})^T$. We also denote a $l_1 \times l_2$ matrix $\mathbf{D}_v = (d_{g_1, g_2 | v})_{l_1 \times l_2}$ where $d_{g_1, g_2 | v} = \eta_{g_1+1, g_2+1 | v} - \eta_{g_1+1, g_2 | v} - \eta_{g_1, g_2+1 | v} + \eta_{g_1, g_2 | v}$. Thus, the estimate of conditional Kendall's tau given in (\ref{eq:tau}) can be rewritten as 
\begin{equation}
\begin{aligned}
\hat{\tau}(v) = 4 \sum_{h_1 = 1}^{l_1} \sum_{h_2 = 1}^{l_2} \eta_{h_1, h_2 | v} \mathbf{a}_{h_1}^T \mathbf{D}_v \mathbf{b}_{h_2} -1.
\end{aligned}
\end{equation}
Furthermore, we can denote two $l_1 \times l_2$ matrices, $\mathbf{H}_v = (\eta_{h_1, h_2 | v})_{l_1 \times l_2}$ and $\mathbf{G}_v = (\mathbf{a}_{h_1}^T \mathbf{D}_v \mathbf{b}_{h_2})_{l_1 \times l_2}$, an as a result, we have
\begin{equation}
\begin{aligned}
\label{eq:tau}
\hat{\tau}(v) = 4 \text{Tr}(\mathbf{H}_v^T \mathbf{G}_v) -1.
\end{aligned}
\end{equation}

Similarly, we are able to rewrite the estimate of conditional Spearman's rho given in (\ref{eq:rho1}). Let us first denote two vectors, $\mathbf{p}_v = (p_{g_1|v})_{l_1}^T$ where $p_{g_1|v} = \eta_{g_1+1, l_2 | v} - \eta_{g_1, l_2 | v}, g_1 = 0, \ldots, l_1-1$ and  $\mathbf{q}_v = (q_{g_2|v})_{l_2}^T$ where $q_{g_2|v} =  \eta_{l_1, g_2+1 | v} - \eta_{l_1, g_2 | v}, g_2 = 0, \ldots, l_2-1$. Then we have 
\begin{equation}
\begin{aligned}
&\hat{\rho}(v) = 12 \sum_{h_1 = 1}^{l_1} \sum_{h_2 = 1}^{l_2}  (\eta_{h_1, h_2 | v} - \eta_{h_1, l_2 | v}     \eta_{l_1, h_2 | v} ) \mathbf{a}_{h_1}^T (\mathbf{p}_v \otimes \mathbf{q}_v) \mathbf{b}_{h_2}. \\
\end{aligned}
\end{equation}
If we further denote two $l_1 \times l_2$ matrices, $\mathbf{R}_v = (r_{h_1, h_2 |v})_{l_1 \times l_2}$ where $r_{h_1, h_2 |v} = \eta_{h_1, h_2 | v} -\eta_{h_1, l_2 | v} \eta_{l_1, h_2 | v}$ and $\mathbf{J}_v = (\mathbf{a}_{h_1}^T (\mathbf{p}_v \otimes \mathbf{q}_v) \mathbf{b}_{h_2})_{l_1 \times l_2}$, then we have
\begin{equation}
\begin{aligned}
\label{eq:rho}
&\hat{\rho}(v) = 12 \text{Tr}(\mathbf{R}_v^T \mathbf{J}_v) \\
\end{aligned}
\end{equation}

By applying the above matrix operations, we are able to obtain very neat expressions of the estimates of conditional dependence measures and the computational efficiency can be improved significantly.

\section{Numerical Illustrations using Simulated Data}
\label{sec:numerical}

We now show the finite-sample performance of the conditional copula estimator $C^{\#}_{x} (u_1, u_2)$. Similar to the simulation setup in \citet{acar2011dependence}, data ${(U_{i1},U_{i2}|X_i), i =1,...,n}$ are generated from the Clayton copula using the package $\tt copula$ in $\tt R$ under the following models: $(U_{i1},U_{i2})|X_i \sim C(u_1,u_2|\theta_i)$, where $\theta_i =\text{exp}(0.8X_i -2)$ and $X_i \sim Unif(0, 3)$. The true copula parameter varies from 0.14 to 1.49 with Spearman's rho ranging from 0.10 to 0.60. The pseudo-observations of the covariate are defined as $V_i \equiv F_{n X}(X_{i}), i= 1, \ldots, n$, where $F_{n X}(x) =  \frac{1}{n+1} \sum_{i=1}^{n} \mathbb{I}(X_{i} \leq x)$. $N= 100$ replicates are drawn from the true copula with sample size $n= 200$. 

Figure \ref{fig:3} shows the contour plots of the Monte Carlo average of the estimated $C^{\#}_{x} (u_1, u_2)$ given $x=0.5$, $x=1$, $x=1.5$, and $x=2$, respectively, across 100 Monte Carlo replicates. The contour plots are drawn based on a $15 \times 15$ equally spaced grid of points in the unit square, meaning that for a given $v$ we need to find $15 + 15 = 30$ roots. Since $F_1$ and $F_2$ are both non-decreasing functions, we can calculate the inverse functions $F_1^{-1}(u_1 | v)$ and $F_2^{-1}(u_2| v)$ by applying the function $\tt uniroot$ in $\tt R$ to equations (\ref{eq:f1}) and (\ref{eq:f2}) for a given value of $v$. The true copula parameters are 0.20 (Spearman's rho equal to 0.09), 0.30 (Spearman's rho equal to 0.20). 0.45 (Spearman's rho equal to 0.27), and 0.67 (Spearman's rho equal to 0.37) for $x=0.5$, $x=1$, $x=1.5$, and $x=2$, respectively.

\begin{figure}
    \centering
           \begin{subfigure}[b]{0.45\textwidth}
        \includegraphics[width=\textwidth]{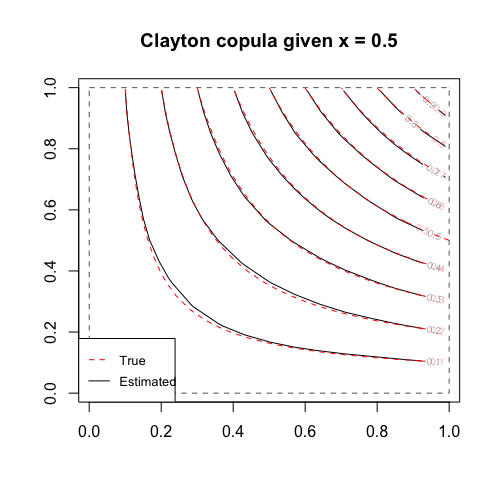}
            \caption{}
    \end{subfigure}
        \begin{subfigure}[b]{0.45\textwidth}
        \includegraphics[width=\textwidth]{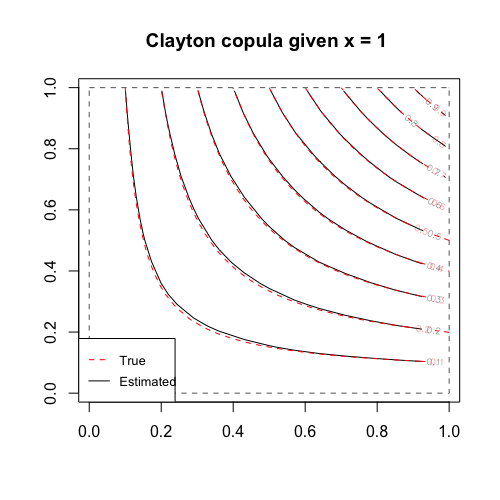}
            \caption{}
    \end{subfigure}
          \begin{subfigure}[b]{0.45\textwidth}
        \includegraphics[width=\textwidth]{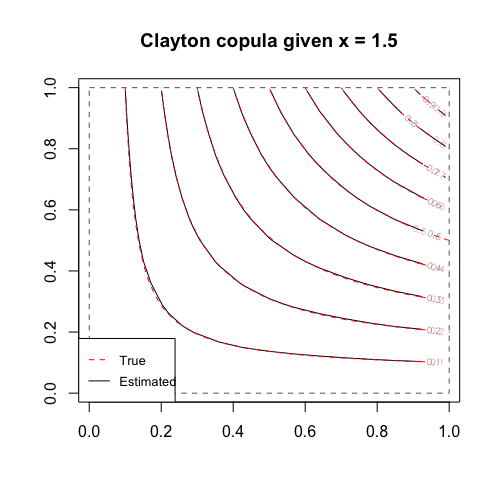}
            \caption{}
    \end{subfigure}
            \begin{subfigure}[b]{0.45\textwidth}
        \includegraphics[width=\textwidth]{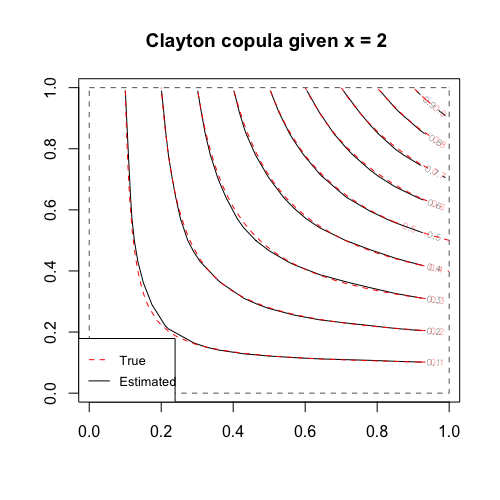}
            \caption{}
    \end{subfigure}
\caption{The contour plots of the Monte Carlo average of the estimated $C^{\#}_{x} (u_1, u_2)$ given $x=0.5$, $x=1$, $x=1.5$, and $x=2$, respectively.}
\label{fig:3}
\end{figure}

It can be observed from the plots that all the estimated contour lines overlap with the true lines at the boundaries, which is a evidence that the conditional copula estimator $C^{\#} (u_1, u_2 | v)$ is a genuine copula with uniform conditional marginal distributions. Moreover, there is almost no bias between the estimated conditional copula averaged over 100 Monte Carlo samples and the true conditional copula across different values of the covariate, illustrating the proposed ECBC-based method works well in estimating conditional copula.

Then we can plot the conditional Kendall's tau and conditional Spearman's rho as given in (\ref{eq:tau}) and (\ref{eq:rho}) as a function of the covariate. The covariate $x$ ranges from 0 to 3 so we compute the dependence measures at seven different values $(0.05, 0.5, 1, 1.5, 2, 2.5, 2.95)$. The following plots show the Monte Carlo average of estimates of dependence measures and the $90\%$ Monte Carlo confidence bands ($5th$ and $95th$ percentiles of
the dependence measure estimates) across 100 Monte Carlo replicates.

Overall, the estimates averaged over 100 Monte Carlo samples seem to be fairly close to the true conditional dependence measures across different values of the covariate. The variance tends to increase and the Monte Carlo average tends to underestimate a little bit when it gets closer to the boundaries of the covariate.

\begin{figure}
    \centering
        \begin{subfigure}[b]{0.55\textwidth}
        \includegraphics[width=\textwidth]{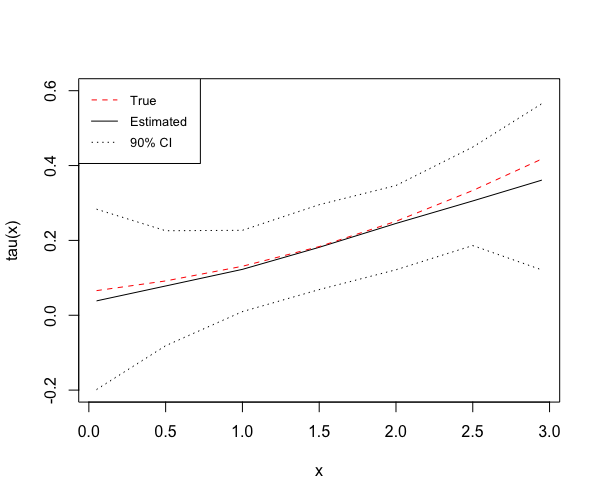}
            \caption{}
    \end{subfigure}
            \begin{subfigure}[b]{0.55\textwidth}
        \includegraphics[width=\textwidth]{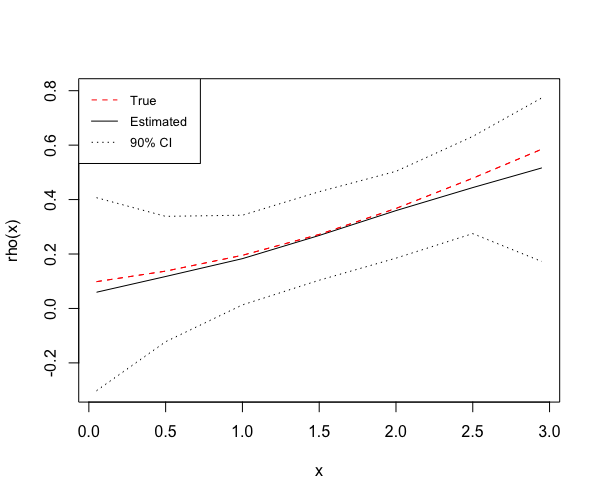}
            \caption{}
    \end{subfigure}
\caption{The plots of the estimated conditional Kendall's tau and conditional Spearman's rho as a function of the covariate.}
\end{figure}

Next, we would like to compare the performance of our proposed nonparametric method to the semiparametric method in \citet{acar2011dependence} through simulation studies. They assume a conditional copula model where the copula function comes from a parametric copula
family and the copula parameter is a function of the covariate. Different copula families, e.g., 
Clayton and Gumbel, were considered and the functional relationship between the copula
parameter and the covariate was estimated using a nonparametric local likelihood approach. 
The severe consequence of misspecified copula model was  investigated in \citet{acar2011dependence} and they proposed a copula selection method based on cross-validated prediction errors. In contrast, the proposed conditional copula estimator is fully nonparametric so there is no need to make any choice of the copula family.

Simulation setups follow \citet{acar2011dependence}. The data ${(U_{i1},U_{i2}|X_i), i =1,...,n}$ are generated from the Clayton copula under the following models: $(U_{i1},U_{i2})|X_i \sim C(u_1,u_2|\theta_i)$, where 
(i): $\theta_i = \text{exp}(0.8X_i -2)$ and $X_i \sim Unif(2, 5)$ ; (ii): $\theta_i = \text{exp}(2-0.3(X_i -2)^2$ and $X_i \sim Unif(2, 5)$. The sample size is $n= 200$.

The comparison can be done numerically by calculating the conditional Kendall's tau and some performance measures, including the integrated square Bias (IBIAS$^2$), integrated Variance (IVAR) and integrated mean square error (IMSE) as given in \citet{acar2011dependence}:
\begin{eqnarray} IBIAS^2 &=& \int_{[2,5]} \big[E[\hat{\tau}_x (x)] - \tau_x(x)\big]^2 d x = 3  \int_{[0,1]} \big[E[\hat{\tau} (v)] - \tau(v)\big]^2 d v,\\
IVAR &=& \int_{[2,5]} E\big[[\hat{\tau}_x (x) - E[\hat{\tau}_x (x)]]^2\big]d x = 3 \int_{[0,1]} E\big[[\hat{\tau} (v) - E[\hat{\tau} (v)]]^2\big]d v,\\
IMSE &=& \int_{[2,5]} E\big[[\hat{\tau}_x (x) - \tau_x(x)]^2\big]d x = 3 \int_{[0,1]} E\big[[\hat{\tau} (v) - \tau(v)]^2\big]d v,
\end{eqnarray}
where the second equality holds because $\tau_x(X) = \tau(F_{ X}(X)) = \tau(V)$ and $X \sim Unif(2, 5)$. We compute Monte Carlo estimates of these performance measures by following the tricks in \citet{segers2017empirical} and compare our proposed method (referred to as "ECBC-based") to the local likelihood method (referred to as "Local") in \citet{acar2011dependence}. The results are shown in Table \ref{tab:comp}.

From the results we can see that when data are generated from the Clayton copula (the underlying true copula), our ECBC-based method outperforms the local likelihood method for the incorrect parametric case (Gumbel) in terms of bias and MSE, although the performance is not as good as the local likelihood method for the correct parametric case (Clayton). Nonetheless, the advantage of the proposed nonparametric method is that we can avoid the adverse impact of misspecified copula and obtain fairly good estimation of conditional copula and conditional dependence measures without having to select the `best' copula model from numerous copula families.

\begin{table}[htbp]
\caption{Comparison of the proposed method (referred to as ``ECBC-based'') to the local likelihood method (referred to as ``Local'') using Monte Carlo estimates of three performance measures, IBIAS$^2$, IVAR and IMSE. Data are generated from the Clayton copula under two different  functional relationships between the copula parameter and the covariate.}
\centering 
\begin{tabular}{ccccc} 
\hline\hline
 \multicolumn{5}{c}{Clayton copula: $\theta =\exp(0.8X -2)$} \\[0.5ex]
\hline
Estimation Method & Parametric Model & IBIAS$^2$ ($\times 10^{-2}$)& IVAR($\times 10^{-2}$) & IMSE ($\times 10^{-2}$)\\[0.5ex]
\hline 
Local & Clayton & 0.017 &0.553 & 0.570 \\ [0.5ex]
\hline 
Local &Gumbel& 3.704& 1.716& 5.389\\[0.5ex]
\hline 
 ECBC-based& N/A & 0.323& 2.569 & 2.892\\[0.5ex]
\hline
\multicolumn{5}{c}{Clayton copula: $\theta =\exp(2 - 0.3(X - 4)^2)$} \\[0.5ex]
\hline
Estimation Method & Parametric Model & IBIAS$^2$ ($\times 10^{-2}$)& IVAR($\times 10^{-2}$) & IMSE ($\times 10^{-2}$)\\[0.5ex]
\hline 
Local & Clayton & 0.040& 0.288 & 0.328 \\ [0.5ex]
\hline 
Local &Gumbel& 4.808 & 1.301 & 6.109\\[0.5ex]
\hline 
 ECBC-based& N/A& 0.855  & 1.876& 2.731\\[0.5ex]
\hline\hline
\end{tabular}
\label{tab:comp}
\end{table}

\section{Real Case Study}
\label{sec:case_study}

We now apply the proposed methodology to a data set of life expectancy at birth of males and females with GDP (in USD) per capita as a covariate for 210 countries or regions. The data are available from the World Factbook 2020 of CIA. Similar data sets were analyzed in \citet{gijbels2011conditional} and \citet{abegaz2012semiparametric}. Life expectancy at birth summarizes the average number of years to be lived in a country while GDP per capita is often considered as an indicator of a country's standard of living. We are interested in the 
dependence between the life expectancy at birth of males and females and would like to see if the strength of dependence is influenced by the GDP per capita. In other words, it is of interest to investigate the dependence between the life expectancy at birth of males ($Y_1$) and females ($Y_2$) conditioning on the covariate $X$, where $X =$ log$_{10}$(GDP) is log$_{10}$ transformation of GDP per capita. 

The pairwise scatterplots of the data are shown in Figure \ref{fig:life}(a), from which we can see that there is strong positive correlation between the 
life expectancy of males (referred to as Male) and females (referred to as Female). Figure \ref{fig:life}(a) also shows that the life expectancy tends to increase with the log$_{10}$ transformation of GDP per capita (referred to as log10.GDP) for both males and females. Before estimating the conditional copula of $(Y_1, Y_2)$ given $X$, we first remove the effect of the covariate $X$ on the marginal distributions of
$Y_1$ and $Y_2$. As a result, the covariate-adjusted pseudo-observations of
$Y_1$ and $Y_2$ (referred to as Male.pseudo and Female.pseudo, respectively) and the pseudo-observations of $X$ (referred to ad log10.GDP.pseudo) are given in \ref{fig:life}(b). 

We then estimate the conditional copula and the conditional dependence of life expectancy at birth of males and females given the covariate $X$. Figure \ref{fig:gdp} shows the estimated conditional Kendall's tau as a function of log$_{10}$(GDP). It can be observed from the plot that the estimate of Kendall's tau decreases from around 0.8 to 0.6 as GDP per capita increases from $10^3 = 1000$ to $10^{4.6} \approx 40000$ USD and it picks up slightly as GDP per capita becomes greater than $40000$ USD. Overall, the dependence between the life expectancy at birth of males and females is relatively larger for countries with lower GDP per capita (less than 10000 USD) and the dependence is relatively smaller for countries with higher GDP per capita (greater than 10000 USD).

\begin{figure}
    \centering
        \begin{subfigure}[b]{0.45\textwidth}
        \includegraphics[width=\textwidth]{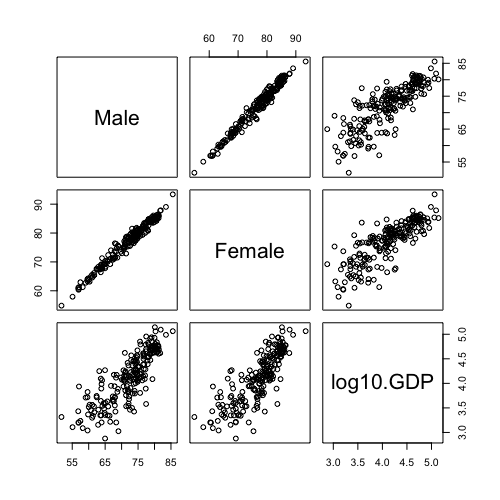}
        \caption{}
    \end{subfigure}
            \begin{subfigure}[b]{0.45\textwidth}
        \includegraphics[width=\textwidth]{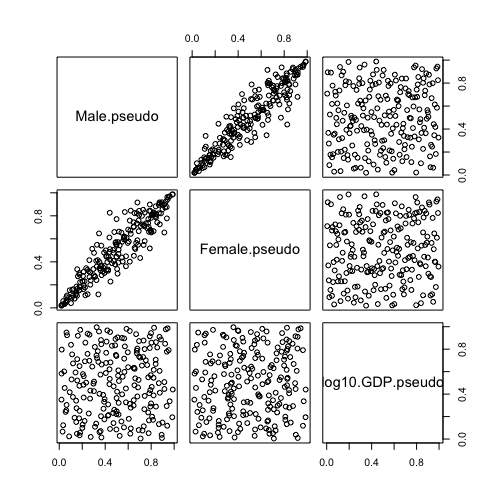}
        \caption{}
    \end{subfigure}
  
\caption{Life expectancy data.}
\label{fig:life}
\end{figure}

\begin{figure}
    \centering
        \begin{subfigure}[b]{0.6\textwidth}
        \includegraphics[width=\textwidth]{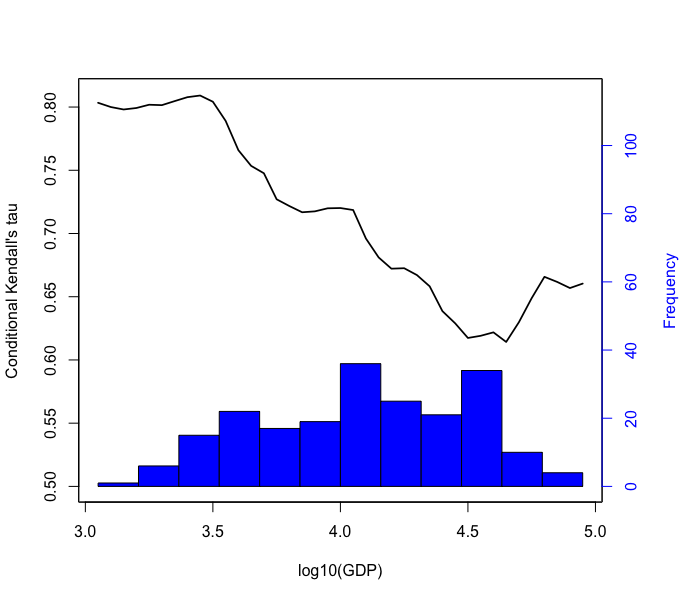}
    \end{subfigure}
  
\caption{Estimated conditional Kendall's tau as a function of log$_{10}$(GDP).}
\label{fig:gdp}
\end{figure}

\section{Concluding Remarks}
\label{sec:conc}

This article provides a nonparametric approach for estimating conditional copulas based on the empirical checkerboard Bernstein copula (ECBC) estimator. The proposed nonparametric method has its own advantages compared to the semiparametric methods as it gets rid of model misspecification by not relying on any selection of copula family and demonstrates a good finite-sample performance. The large-sample consistency of the proposed ECBC-based conditional copula estimator is also presented. In addition, we derive closed-form nonparametric estimates of the conditional dependence measures from the proposed  estimator.

Due to the complexity in modelling and inference caused by the dependence of conditional copula on the covariates, it is quite common in practice, particularly for vine copulas, to assume that the dependence structure is not influenced by the value of covariates, which is referred to as `simplifying assumption' (e.g. see \citet{haff2010simplified}, \citet{acar2012beyond}, \citet{stoeber2013simplified}, \citet{nagler2016evading} and \citet{schellhase2018estimating}). In the literature, there have been some available tests of the simplifying assumption; see \citet{acar2013statistical}, \citet{gijbels2017score}, \citet{gijbels2017nonparametric}, \citet{derumigny2017tests} and \citet{kurz2017testing}, etc. Our proposed ECBC-based conditional copula estimator can be useful for constructing new tests of the simplifying assumption. We have shown the framework of obtaining a general estimate of the conditional copula that is allowed to vary with the value of covariates. It is also straightforward to obtain an estimate satisfying the simplifying assumption based on the covariate-adjusted pseudo-observations again using ECBC estimator. Therefore, it could be possible to build test statistics based on some discrepancy criteria like Kolmogorov-Smirnov type, Anderson-Darling type, etc., where the distributions of such test statistics could be approximated by bootstrap schemes.

Another interesting topic for future work would be extending the estimation framework to high-dimensional conditional copula. We can perhaps first use some dimension reduction methods like principal component analysis (PCA) and then develop copula models based on the lower dimensional principal components of the covariates.

\bibliographystyle{chicago}
\bibliography{main}

\newpage
\section*{Appendix}

\noindent \underline{Proof of Theorem \ref{thm:consist}}

\begin{proof}
We denote
\begin{equation}
P_{m,k} (v) = {m \choose k} v^k (1-v)^{m-k}.
\end{equation}
Then we can rewrite the Bernstein copula as
\begin{equation}
\begin{aligned}
B (C; u_1, u_2, v) = \sum_{h_1 = 0}^{l_1} \sum_{h_2 = 0}^{l_2} \sum_{k = 0}^{m} C\left(\frac{h_1}{l_1}, \frac{h_2}{l_2}, \frac{k}{m}\right)   P_{l_1,h_1} (u_1) P_{l_2,h_2} (u_2) P_{m,k} (v),
\end{aligned}
\end{equation}
and the ECBC copula estimator as
\begin{equation}
\begin{aligned}
B (C_n^{\#}; u_1, u_2, v) = \sum_{h_1 = 0}^{l_1} \sum_{h_2 = 0}^{l_2} \sum_{k = 0}^{m} C^{\#}_n\left(\frac{h_1}{l_1}, \frac{h_2}{l_2}, \frac{k}{m}\right)  P_{l_1,h_1} (u_1) P_{l_2,h_2} (u_2) P_{m,k} (v),
\end{aligned}
\end{equation}
where $C^{\#}_n$ is the empirical checkerboard copula. Thus the partial derivative of 3-dimensional ECBC $B (C_n^{\#}; u_1, u_2, v)$ with respect to $v$ takes the form of 
\begin{equation}
\begin{aligned}
& C^{\# (1)} (u_1, u_2 | v) \equiv \frac{\partial B (C_n^{\#}; u_1, u_2, v)}{\partial v } \\
&= \sum_{h_1 = 0}^{l_1} \sum_{h_2 = 0}^{l_2} \sum_{k = 0}^{m}  C^{\#}_n\left(\frac{h_1}{l_1}, \frac{h_2}{l_2}, \frac{k}{m}\right)  P_{l_1,h_1} (u_1) P_{l_2,h_2} (u_2) P_{m,k}^{'} (v),
\end{aligned}
\end{equation}
where $P_{m,k}^{'} (v)$ is the derivative of $ P_{m,k} (v)$ with respect to $v$.

Let us denote the partial derivative of the Bernstein copula $B (C; u_1, u_2, v)$ with respect to $v$ as
\begin{equation}
\begin{aligned}
& C^{ (1)} (u_1, u_2 | v) \equiv \frac{\partial B (C; u_1, u_2, v)}{\partial v } \\
&= \sum_{h_1 = 0}^{l_1} \sum_{h_2 = 0}^{l_2} \sum_{k = 0}^{m}  C\left(\frac{h_1}{l_1}, \frac{h_2}{l_2}, \frac{k}{m}\right)  P_{l_1,h_1} (u_1) P_{l_2,h_2} (u_2) P_{m,k}^{'} (v),\\
\end{aligned}
\end{equation}
and the partial derivative of the empirical Bernstein copula $B (C_n; u_1, u_2, v)$ with respect to $v$ as
\begin{equation}
\begin{aligned}
& C^{ (1)}_n (u_1, u_2 | v) \equiv \frac{\partial B (C_n; u_1, u_2, v)}{\partial v } \\
&= \sum_{h_1 = 0}^{l_1} \sum_{h_2 = 0}^{l_2} \sum_{k = 0}^{m}  C_n \left(\frac{h_1}{l_1}, \frac{h_2}{l_2}, \frac{k}{m}\right)  P_{l_1,h_1} (u_1) P_{l_2,h_2} (u_2) P_{m,k}^{'} (v).\\
\end{aligned}
\end{equation}
Using the triangle inequality we have
\begin{equation*}
\begin{aligned}
||  C^{\#(1)}  - C_v||(v) &\leq || C^{\# (1)}  - C^{ (1)} ||(v) +  ||C^{ (1)}  - C_v|(v)  \\
&\leq || C^{\# (1)}  - C^{ (1)}_n ||(v) + ||C^{ (1)}_n  - C^{ (1)}|(v) + ||C^{ (1)}  - C_v||(v).
\end{aligned}
\end{equation*}
First, we can show that
\begin{equation*}
\begin{aligned}
& || C^{\# (1)}  - C^{(1)}_n ||(v) \\
&=  ||\sum_{h_1 = 0}^{l_1} \sum_{h_2 = 0}^{l_2} \sum_{k = 0}^{m} \left( C^{\#}_n\left(\frac{h_1}{l_1}, \frac{h_2}{l_2}, \frac{k}{m}\right)-   C_n\left(\frac{h_1}{l_1}, \frac{h_2}{l_2}, \frac{k}{m}\right)\right)  P_{l_1,h_1} (u_1) P_{l_2,h_2} (u_2) P_{m,k}^{'} (v) || (v)\\
& \leq  \underset{0 \leq h_1 \leq l_1, 0 \leq h_2 \leq l_2, 0 \leq k \leq m-1}{\max} \left| C_n^{\#}\left(\frac{h_1}{l_1}, \frac{h_2}{l_2}, \frac{k+1}{m}\right) -  C_n\left(\frac{h_1}{l_1}, \frac{h_2}{l_2}, \frac{k+1}{m}\right)\right| \\
&\sum_{h_1 = 0}^{l_1} \sum_{h_2 = 0}^{l_2} \sum_{k = 0}^{m} | P_{l_1,h_1} (u_1) | | P_{l_2,h_2} (u_2)| | P_{m,k}^{'} (v)  | \\
& \leq  \underset{0 \leq h_1 \leq l_1, 0 \leq h_2 \leq l_2, 0 \leq k \leq m-1}{\max} \left| C_n^{\#}\left(\frac{h_1}{l_1}, \frac{h_2}{l_2}, \frac{k+1}{m}\right) -  C_n\left(\frac{h_1}{l_1}, \frac{h_2}{l_2}, \frac{k+1}{m}\right)\right| \sum_{k = 0}^{m} | P_{m,k}^{'} (v)  |. \\
\end{aligned}
\end{equation*}
In above the second inequality follows from the fact that since ${l_j \choose h_j } u_j^{l_j} (1 - u_j)^{l_j - h_j}, l_j = 0,1,\ldots,h_j$, $j=1,2$ are binomial probabilities, $\sum_{h_j=0}^{l_j}{l_j \choose h_j } u_j^{h_j} (1 - u_j)^{l_j - h_j}=1$ for any $u_j\in [0, 1]$, $j =1, 2$.
Under the assumption that the marginal CDFs are continuous, it follows from the Remark 2 in \citet{genest2017asymptotic} that for d-dimensional copula
\begin{equation*}
|| C_n^{\#} - C_n|| \leq \frac{3}{n},
\end{equation*}
and from Lemma 1 in \citet{janssen2014note} it follows that for any fixed $0 < v < 1$,
\begin{equation*}
\sum_{k = 0}^{m} | P_{m,k}^{'} (v) | \sim \sqrt{\frac{2}{\pi}} \frac{m^{1/2}}{\sqrt{v(1-v)}} = O(m^{1/2}) \; \; \text{as} \; \; m \rightarrow \infty.
\end{equation*}
Thus,  for any fixed $0 < v < 1$ we have
\begin{equation*}
\begin{aligned}
& || C^{\# (1)}  - C^{ (1)}_n ||(v) \\
& \leq  \underset{0 \leq h_1 \leq l_1, 0 \leq h_2 \leq l_2, 0 \leq k \leq m-1}{\max} \left| C_n^{\#}\left(\frac{h_1}{l_1}, \frac{h_2}{l_2}, \frac{k+1}{m}\right) -  C_n\left(\frac{h_1}{l_1}, \frac{h_2}{l_2}, \frac{k+1}{m}\right)\right| \sum_{k = 0}^{m} | P_{m,k}^{'} (v)  | \\
& \leq || C_n^{\#} - C_n||\sum_{k = 0}^{m} | P_{m,k}^{'} (v) | 
= O(m^{1/2} n^{-1}).
\end{aligned}
\end{equation*}
Next, we can use a similar technique to show that 
\begin{equation*}
\begin{aligned}
&||C^{ (1)}_n  - C^{ (1)}|(v) \\
& \leq  \underset{0 \leq h_1 \leq l_1, 0 \leq h_2 \leq l_2, 0 \leq k \leq m-1}{\max} \left| C_n\left(\frac{h_1}{l_1}, \frac{h_2}{l_2}, \frac{k+1}{m}\right) -  C\left(\frac{h_1}{l_1}, \frac{h_2}{l_2}, \frac{k+1}{m}\right)\right| \sum_{k = 0}^{m} | P_{m,k}^{'} (v)  | \\
\end{aligned}
\end{equation*}
By using the  Lemma 1 in \citet{janssen2012large} and equation (3) in \citet{kiriliouk2019some}, for d-dimensional copula we obtain
\begin{equation*}
\begin{aligned}
|| C_n - C|| & \leq \frac{3}{n} + O(n^{-1/2}(\log \log n)^{1/2})\;\;\;a.s.\\
&= O(n^{-1/2}(\log \log n)^{1/2})\;\;\; a.s..
\end{aligned}
\end{equation*}
Thus it follows that for any fixed $0 < v < 1$,
\begin{equation*}
||C^{ (1)}_n  - C^{ (1)}||(v) =  O(m^{1/2}n^{-1/2}(\log \log n)^{1/2}),\;\;\; a.s..
\end{equation*}
Hence, for any fixed $0 < v < 1$ we have
\begin{equation*}
|| C^{\# (1)}  - C^{ (1)} ||(v) \leq || C^{\# (1)}  - C^{ (1)}_n ||(v) + ||C^{ (1)}_n  - C^{ (1)}|(v) = O(m^{1/2}n^{-1/2}(\log \log n)^{1/2}),\;\;\; a.s..
\end{equation*}
Next, by mean value theorem there exists $\frac{k}{m} < \xi_k < \frac{k+1}{m}$ s.t.
\begin{equation*}
\begin{aligned}
& C^{ (1)} (u_1, u_2 | v)\\
&= \sum_{h_1 = 0}^{l_1} \sum_{h_2 = 0}^{l_2} \sum_{k = 0}^{m-1} m \left( C\left(\frac{h_1}{l_1}, \frac{h_2}{l_2}, \frac{k+1}{m}\right) -  C\left(\frac{h_1}{l_1}, \frac{h_2}{l_2}, \frac{k}{m}\right) \right) P_{l_1,h_1} (u_1) P_{l_2,h_2} (u_2) P_{m-1,k}(v),\\
&= \sum_{h_1 = 0}^{l_1} \sum_{h_2 = 0}^{l_2} \sum_{k = 0}^{m-1} C_v\left(\frac{h_1}{l_1}, \frac{h_2}{l_2}\big| \xi_k \right) P_{l_1,h_1} (u_1) P_{l_2,h_2} (u_2) P_{m-1,k}(v),\\
&= \sum_{h_1 = 0}^{l_1} \sum_{h_2 = 0}^{l_2} \sum_{k = 0}^{m-1} \left(C_v\left(\frac{h_1}{l_1}, \frac{h_2}{l_2}\big| \xi_k \right) - C_v\left(\frac{h_1}{l_1}, \frac{h_2}{l_2}\big| \frac{k}{m-1} \right) +C_v\left(\frac{h_1}{l_1}, \frac{h_2}{l_2}\big|  \frac{k}{m-1} \right)\right) \\
& P_{l_1,h_1} (u_1) P_{l_2,h_2} (u_2) P_{m-1,k}(v).\\ 
\end{aligned}
\end{equation*}
Notice that 
\begin{equation*}
\begin{aligned}
 \frac{k}{m} - \frac{k+1}{m} <\frac{k}{m} - \frac{k}{m-1}<  \xi_k -  \frac{k}{m-1} <  \frac{k+1}{m} -  \frac{k}{m-1} < \frac{k+1}{m} - \frac{k}{m},
\end{aligned}
\end{equation*}
which means that 
\begin{equation*}
\begin{aligned}
\left| \xi_k -  \frac{k}{m-1}\right| <  \frac{1}{m}.
\end{aligned}
\end{equation*}
If $C_v(u_1, u_2 | v) = \frac{\partial C(u_1, u_2, v)}{\partial v}$ is Lipschitz continuous on $[0,1]^3$, then there exists a Lipschitz constant $L$ s.t.
\begin{equation*}
\begin{aligned}
\left|C_v\left(\frac{h_1}{l_1}, \frac{h_2}{l_2}\big| \xi_k \right) - C_v\left(\frac{h_1}{l_1}, \frac{h_2}{l_2} \big| \frac{k}{m-1} \right)\right| \leq L \left| \xi_k -  \frac{k}{m-1} \right|  \leq \frac{L}{m},
\end{aligned}
\end{equation*}
and based on Lemma 3.2 in \citet{segers2017empirical} we also have 
\begin{equation*}
\begin{aligned}
&|| \sum_{h_1 = 0}^{l_1} \sum_{h_2 = 0}^{l_2} \sum_{k = 0}^{m-1} C_v\left(\frac{h_1}{l_1}, \frac{h_2}{l_2} \big|  \frac{k}{m-1} \right)P_{l_1,h_1} (u_1) P_{l_2,h_2} (u_2) P_{m-1,k}(v)
- C_v|| \\
&\leq L \left(\frac{1}{2 \sqrt{l_1}} + \frac{1}{2 \sqrt{l_2}} + \frac{1}{2 \sqrt{m-1}} \right).
\end{aligned}
\end{equation*}
Thus 
\begin{equation*}
\begin{aligned}
&|| C^{(1)} - C_v ||(v) \\
&\leq || \sum_{h_1 = 0}^{l_1} \sum_{h_2 = 0}^{l_2} \sum_{k = 0}^{m-1} \left(C_v\left(\frac{h_1}{l_1}, \frac{h_2}{l_2} \big| \xi_k \right) - C_v\left(\frac{h_1}{l_1}, \frac{h_2}{l_2} \big| \frac{k}{m-1} \right)\right)  P_{l_1,h_1} (u_1) P_{l_2,h_2} (u_2) P_{m-1,k}(v)||\\ 
&+ || \sum_{h_1 = 0}^{l_1} \sum_{h_2 = 0}^{l_2} \sum_{k = 0}^{m-1} C_v\left(\frac{h_1}{l_1}, \frac{h_2}{l_2}\big|  \frac{k}{m-1} \right)P_{l_1,h_1} (u_1) P_{l_2,h_2} (u_2) P_{m-1,k}(v) - C_v|| \\
& \leq \frac{L}{m} + L \left(\frac{1}{2 \sqrt{l_1}} + \frac{1}{2 \sqrt{l_2}} + \frac{1}{2 \sqrt{m-1}} \right).
\end{aligned}
\end{equation*}
Finally, for any fixed $0 < v < 1$ we obtain
\begin{equation*}
\begin{aligned}
||  C^{\# (1)}  - C_v||(v)  &\leq ||  C^{\# (1)}  - C^{(1)}||(v) + || C^{(1)} - C_v ||(v) \\
&\leq \frac{L}{m} + L \left(\frac{1}{2 \sqrt{l_1}} + \frac{1}{2 \sqrt{l_2}} + \frac{1}{2 \sqrt{m-1}} \right) + O(m^{1/2}n^{-1/2}(\log \log n)^{1/2}),\;\;\; a.s..
\end{aligned}
\end{equation*}
The empirical prior of the degrees $m$, $l_1$, and $l_2$ are given as
\begin{equation*}
m |\alpha \sim Poisson(n^{\alpha}) + 2\;\;\mbox{and}\;\;
\alpha \sim Unif\Big(\frac{1}{3}, \frac{2}{3}\Big), \; \;
\end{equation*}
\begin{equation*}
l_j |\alpha_j \sim Poisson(n^{\alpha_j}) + 1\;\;\mbox{and}\;\;
\alpha_j \sim Unif\Big(\frac{1}{3}, \frac{2}{3}\Big) \; \;j= 1,2.
\end{equation*}
Notice that $\Pr({1\over 3} \leq \alpha \leq {2\over 3})=1$ and $\Pr({1\over 3} \leq \alpha_j \leq {2\over 3})=1, j=1,2$, then $E(m^{1/2}n^{-1/2}(\log \log n)^{1/2}) $ $ \leq  n^{1/3} n^{-1/2}(\log \log n)^{1/2} \rightarrow 0$ as $n \rightarrow \infty$. In the proof of Theorem 3.1 it has been shown $E\left(\sqrt{\frac{1}{l_j}}\big|\alpha_j\right)\leq \sqrt{{1-e^{-n^{\alpha_j}}\over n^{\alpha_j}}}\rightarrow 0, j = 1, 2$, $E\left(\sqrt{\frac{1}{m-1}}\big|\alpha\right)\leq \sqrt{{1-e^{-n^{\alpha}}\over n^{\alpha}}}\rightarrow 0$ as $n \rightarrow \infty$ and $E\left(\frac{1}{m}\big|\alpha\right) \leq {1-e^{-n^{\alpha}}\over n^{\alpha}}\rightarrow 0$ as $n \rightarrow \infty$. Thus, taking expectation with respect to the prior distributions of $l_1$, $l_2$ and $m$ as given for ECBC, it follows that 
\begin{equation*}
\begin{aligned}
&E (||  C^{\# (1)}  - C_v||(v))\\
&\leq E\left(\frac{L}{m}\right) + E\left(L\left(\frac{1}{2 \sqrt{l_1}} + \frac{1}{2 \sqrt{l_2}} + \frac{1}{2 \sqrt{m-1}} \right)\right) +  E\left(O(m^{1/2}n^{-1/2}(\log \log n)^{1/2})\right)\\
& \rightarrow 0 \; as \; n \rightarrow \infty\;\; a.s.\\
\end{aligned}
\end{equation*}

Since $C(u_1, 1 | v) = u_1$ and $C(1, u_2 | v) = u_2$ and $C^{\# (1)} (u_1, u_2 |v)$ converges to $ C_v(u_1, u_2 | v)$ uniformly on $[0,1]^2$ as $n \rightarrow  \infty$ for any fixed $0 < v < 1$, then we have
\begin{equation*}
\begin{aligned}
&E\left(||F_1(u_1 | v) - u_1||(v)\right) \equiv \displaystyle E \left(\sup_{u_1\in [0, 1]}| F_1(u_1 | v) - u_1|\right) \\
&= E\left(||C^{\# (1)} (u_1, 1 | v) - C(u_1, 1 | v)||(v)\right) \leq E\left( || C^{\# (1)}  - C^{ (1)} ||(v)\right) \stackrel{\text{a.s.}}{\rightarrow} 0,
\end{aligned}
\end{equation*}
and
\begin{equation*}
\begin{aligned}
&E\left(||F_2(u_2 | v) - u_2||(v)\right) \equiv \displaystyle E \left(\sup_{u_2\in [0, 1]}| F_2(u_2 | v) - u_2| \right) \\
&= E\left(||C^{\# (1)} (1, u_2 | v) - C(1, u_2 | v)||(v) \right)\leq E\left(|| C^{\# (1)}  - C^{ (1)} ||(v)\right) \stackrel{\text{a.s.}}{\rightarrow} 0.
\end{aligned}
\end{equation*}
For any fixed $0 < v < 1$, $F_1$ and $F_2$ are non-decreasing functions, so $F_1^{-1}(u_1 | v)  \stackrel{\text{a.s.}}{\rightarrow} u_1$ and $F_2^{-1}(u_2 | v) \stackrel{\text{a.s.}}{\rightarrow} u_2$, Thus we obtain the uniform convergence of
\begin{equation*}
\begin{aligned}
E(||  C^{\#}  - C_v||(v) ) = E(||  C^{\#(1)} (F_1^{-1}(u_1 |v ), F_2^{-1}(u_2| v) | v)  - C_v||(v) ) \stackrel{\text{a.s.}}{\rightarrow} 0
\end{aligned}
\end{equation*}
as $n \rightarrow  \infty$ for any fixed $0 < v < 1$.

\end{proof}

\end{document}